\documentclass{aastex}

\slugcomment{Not to appear in Nonlearned J., 45.}

\shorttitle{Uranus irregular satellites.}
\shortauthors{M.Maris et al}

\begin{document}

\title{Multicolor Photometry of the Uranus Irregular Satellites
Sycorax and Caliban\thanks{Based on observations carried out at ESO La Silla
(Chile)}}

\author{Michele Maris}
\affil{Osservatorio Astronomico di Trieste - Via G.B. Tiepolo, 11 - I 34131 - Trieste - Italy}
\email{maris@ts.astro.it}

\author{Giovanni Carraro}
\affil{ Dipartimento di Astronomia, Universit\'a di Padova - Vicolo dell'Osservatorio, 5 - I 35122 - Padova - Italy}
\email{carraro@pd.astro.it}

\author{Gabriele Cremonese}
\affil{ Osservatorio Astronomico di Padova - Vicolo
              dell'Osservatorio, 5 - I 35122 - Padova - Italy}
\email{cremonese@pd.astro.it}

\and

\author{Marco Fulle}
\affil{Osservatorio Astronomico di Trieste - Via G.B. Tiepolo, 11 - I 34131 - Trieste - Italy}
\email{fulle@ts.astro.it}

\begin{abstract}
 We report on accurate BVRI photometry for the two Uranus irregular
 satellites Sycorax and Caliban. We derive colours, showing that
 Sycorax is bluer than Caliban.  Our data allows us to detect a
 significant variability in the Caliban's light-curve, which suggests an
 estimated period of about 3 hours. Despite it is the brighter of
 the two bodies, Sycorax does not display a strong
 statistically significant variability. However our data seem to suggest
 a period of about 4 hours.
\end{abstract}


\section{Introduction}
A couple of new Uranus satellites, named Sycorax (S/1997 U1) and
Caliban (S/1997 U2), with an orbital semimajor axis of
253 and 305 uranian radii, respectively,  were discovered in 1997 by
Gladman et.al. \cite{Gladman:etal:1998}. This solved the apparent Uranus {\em
peculiarity} known up to that time to be the only giant planet in
the Solar System without irregular satellites, despite accurate
search carried out in the past
(Christie \cite{Christie:1930},
 Kuiper \cite{Kuiper:1961},
 Smith \cite{Smith:1984},
 Cruikshank \& Brown \cite{Cruikshank:Brown:1986}).

Giant gaseous planets are characterized by a complex system of
dust rings and satellites. From the point of view of orbital
dynamics, the satellites of the giant gaseous planets belong to two
classes: regular and irregular. The former are characterized by
orbits with a small eccentricity, very close to the planet
equatorial plane and always show a prograde motion. The latter
follow orbits which may have a large ellipticity, semimajor axis, and
inclination. Moreover they may follow both prograde and retrograde
motions. According to Pollack et al. \cite{Pollack:etal:1979} the
 two classes of satellites suggest a quite different evolution. Regular
satellites are supposed to be born in the same subnebula from
which the planet originated. On the other hand irregular
satellites might be planetesimals  felt inside the planet
subnebula by gas drag just before the subnebula collapse.
Eventually the captured planetesimals were fragmented by dynamical
pressure due to the gas drag and were dispersed by
collisions with other objects already present in the subnebula
(Pollack et al. \cite{Pollack:etal:1979}).
Following this scheme, it is evident
that the study of irregular satellites is important in the context
of the Solar System origin. In particular, it could be interesting
to compare the newly discovered uranian irregular satellites
with other classes of minor bodies in the outer Solar System,
i.e.  Kuiper Belt and Centaurus objects.

The faintness of the new satellites (Gladman et al. \cite{Gladman:etal:1998}
report $R_{Sycorax}\approx 20.4$\ mag, $R_{Caliban}\approx 21.9$\
mag) made it difficult to determine their photometric properties.
Colors are reported by Gladman et al. \cite{Gladman:etal:1998} with 0.1 mag
accuracy, suggesting that Sycorax and Caliban are reddened with
respect to the Sun, in contrast with Uranus and its regular
satellites. Moreover, the low photometric accuracy ($\sim
0.1$\ mag) prevented them from obtaining a light-curve and
hence an estimate of the rotation period for both satellites.

To improve the present knowledge, Sycorax and Caliban has been
observed with the NTT ESO telescope. We obtained magnitudes in the
B, V, R, I bands with accuracy of 0.03 mag and we obtain color and
light-curves.

The paper is organized as follows: section 2 describes the data
acquisition and reduction, section 3 the colors, and section 4 the
light curve. Final remarks and conclusions are given in
section 5.

\section{Data acquisition and reduction}
Observations were conducted at La Silla on 1999 October 8 and 9,
using the Tektronix 2024$\times$ 2024 pixel CCD $\#$ 36 mounted in
the red EMMI arm  of the 3.6-m ESO NTT telescope. The first night
was photometric with an average seeing of 1.0 arcsec, whereas the
second one was not photometric. The scale on the chip is 0.27
arcsec per pixel, and the array covers about 9$\times$9
arcmin$^{2}$ in the sky. Details of the observing run for the two
satellites are given in Tables \ref{tab:Sycorax:Journal} and
\ref{tab:Caliban:Journal}.

Pre-processing, which includes bias and flat-field corrections,
has been done by using standard routines in the ESO MIDAS package.

Instrumental magnitudes have been extracted at the Padova University
using the DAOPHOT and accompanying ALLSTAR package
(Stetson \cite{Stetson:1991}) in the MIDAS environment. The errors in
Tables \ref{tab:Sycorax:Journal} and \ref{tab:Caliban:Journal} are
assumed to be normally distributed and are 1$\sigma$.

The instrumental $b$, $v$, $r$ and $i$ have been transformed into
standard Jonhson $B$, $V$ and Cousin  $R$ and $I$ magnitudes using
fitting coefficients (colour term and zero point) derived from
observations of the standard field T-Phoenix and MarkA stars from
Landolt (1992) at the beginning and the end of the night, after
including exposure time normalization and airmass corrections.
Aperture correction has also been applied. The transformations are
given by the following equations:

\begin{eqnarray}\label{eq:Colour}
 (B-b) &=& -0.050 \times (B-V) - 0.414\label{eq:Colour:B-b}\\
 (V-v) &=& 0.010 \times (B-V) + 0.030\\
 (B-V) &=& 0.994 \times (b-v) - 0.437\\
 (V-v) &=& 0.010 \times (V-I) + 0.029\\
 (V-I) &=& 1.003  \times (v-i) + 0.551\\
 (V-v) &=& 0.020 \times (V-R) - 0.029\\
 (R-r) &=&  0.01 \times (V-R)  +0.157\\
 (V-R) &=& 1.010 \times (v-r) - 0.101
\end{eqnarray}

\noindent and are plotted in Fig.~1. The errors affecting this
calibration are expected to be of the order of 0.03 mag. All the
magnitudes obtained in the second night were translated into the
first night through the comparison of a  group of  common field
stars in the same frames of the satellites.

\section{Colors}

The observations allowed us to obtain the colors of both
satellites with an excellent accuracy and we reported them in
Table \ref{tab:colours}\ together with corresponding error bars. The colors are
almost similar for both satellites, even if Sycorax appears to be
bluer than Caliban looking at (B-V), and we can make a comparison
with other minor bodies of the Solar System. They clearly appear
less red than most of the Kuiper Belt objects as it can be seen on
the histogram of V-R reported by Luu \& Jewit \cite{Luu:Jewit:1996}, while
there is an interesting similarity with some Centaurs, 1995 GO,
1997 CU$_{26}$ and 1995 DW$_2$, as it is well shown in the Table V
of Davies et al. \cite{Davies:etal:1998}.

However, it is difficult to provide even a rough interpretation of
the colors measured for Sycorax and Caliban. Looking at the
histograms of Luu \& Jewitt \cite{Luu:Jewit:1996} the V-R
values of the satellites could be considered as the bluest Kuiper
Belt objects or the reddest Near Earth Asteroids, not neglecting
that 0.5 is the value found for most of the comet nuclei and the
Trojans. The small number of Centaurus observed up to now shows a
broad range of V-R values and it might be too much easy to
associate Sycorax and Caliban to this group, even if the
heliocentric distance of Uranus is closer to their semi-major
axis.

\section{Light Curves}\label{sec:Light:Curves}

Light-curves have been obtained in the R band with exposure times
of 600 - 800 sec for Caliban, 80 - 200 sec for Sycorax. The time
dependent part of the light curves together with a sinusoidal fit
described below, are plotted in Fig. \ref{fig2}.

The first night only has been photometric, but for both
satellites the first night alone is not sufficient to
say anything of conclusive about the time variability
even applying simple models. So the data of
the second night have been translated to the first night by
matching the magnitude of common stars in the frames.
The chosen stars do not exhibit significant brightness variations,
however this procedure may left some residual systematic
calibration error within a $1\sigma$\ level ($\approx 0.03$\ mag) equivalent
to a shift in the zero point of the magnitude scale for the second night
relatively to the scale of the first night.
Since data generally are not evenly distributed about the mean, it is not possible to
remove this shift, subtracting from the data of each night their average.
As a consequence, a possible small shift in the zero point of the magnitude scale
between different nights must be accounted for by the fit itself adding a
further degree of freedom to the model. So the fit is:
\begin{equation}\label{eq:sinfit}
    R(t) = R_0 + \Delta \, \mathrm{U}(t - t_0) + A \, \sin\left( \, \frac{2\pi}{P} (t-t_0) + \phi \, \right)
\end{equation}

\noindent whose four free parameters are: the average
magnitude $R_0$, the amplitude $A$, the period $P$, and the phase
$\phi$. The origin of time $t_0$\ is assumed to be October 8, 2000 at
UT 00:00.
The step function $\mathrm{U}$\ is null during the
first night and +1 during the second one. The free parameter
$\Delta$\ accounts for the possible shift in the zero point between
the two nights.
If this is the case, $\Delta$\ will assume values significantly different from zero.
The fit is performed twice by weighted least squares.
The first time imposing $\Delta \equiv 0$\ and the second time
leaving it as a fifth free parameter. The results of the two fits
are then compared. As shown below Sycorax data requires a significant
shift between the first and the second night, while Caliban does not.
In principle more sophisticated methods could be used,
but our limited data set does not justify their application.
As a consequences the lightcurve parameters,
and particularly their periods, shall be considered just as indicative estimates
rather than firm results.

In the first night Caliban does not display a large
variation ($\Delta R = 0.057 \pm 0.032$ mag, i.e. less than 2
$\sigma$), while in the second one it shows $\Delta R =
0.237\pm0.045$\ mag, corresponding to a $\approx5\sigma$\ level. The
first concern was to verify whether such variability may be
explained by random fluctuations due to noise or not. The
$\chi^2$\ test rejected the hypothesis of random fluctuations. In
fact taking into account the data of both nights, the
significativity level for this hypothesis is less than $0.005\%$.
In the case of Caliban no significant shift is required between
the magnitude zero points of first and second nights.
Leaving the shift $\Delta$\ as a free parameter does not improve
the fit, but it reduces the number of degrees of freedom and so
the significativity level for the fit. Then we imposed
$\Delta \equiv 0$\ obtaining the best fit for
 $P = 2.6624 \pm 0.0130$\ hours,
 $\phi = 4.2607 \pm 0.1637$\ rad,
 $A = 0.1169  \pm 0.0102$\ mag,
 $R_0 = 21.9128 \pm 0.0112$\ mag,
 with
 $\chi^2 = 2.6331$\
 equivalent to a significativity level
 $SL = 75.63\%$.

Indeed most of the information in this estimate is based on the data of the second
night. As a comparison the best fit obtained considering the second
night data only is obtained for
 $P = 2.7011 \pm 0.0093$\ hours,
 $\phi = 5.1269 \pm 0.1061$\ rad,
 $A = 0.1342 \pm 0.0128$\ mag,
 $R_0 = 21.9126 \pm 0.0120$\ mag,
 with
 $\chi^2 = 0.1739$\
 equivalent to a significativity level
 $SL = 98.17\%$.
It would be noted how a better fit is obtained for the second night data than for the
full data set, the worst behaving point being the last of the first night.
If this point is removed we obtain
 $P = 2.6678  \pm 0.0119$\ hours,
 $\phi = 4.4204\pm 0.1428$\ rad,
 $A = 0.1268 \pm 0.0171$\ mag,
 $R_0 = 21.9037 \pm 0.0120$\ mag,
 with
 $\chi^2 = 0.6860$\
 equivalent to a significativity level
 $SL = 98.37\%$.
This fact suggests either a residual mismatch in the zero point calibration between
the two nights or
that the light curve is not properly represented by a sinusoidal time dependence.
However it is not possible to discriminate between these two possibilities from the
data as none of the related CCD frames display evident peculiarities.
So the
difference between the two results will be regarded as an estimate of
the systematic errors in the light curve parameters determination.

In conclusion our best estimates are:
 $P    =  2.66   ^{+0.04}_{-0.00} \pm 0.01$\ hours,
 $\phi =  4.26   ^{+0.87}_{-0.00} \pm 0.16$\ rad,
 $A    =  0.134  ^{+0.000}_{-0.008} \pm 0.010$\ mag,
 $R_0  =  21.913 ^{+0.000}_{-0.000} \pm 0.011$\ mag.
Where the first error refers to the systematic error and the second
to the random ($1\sigma$) error.

For Sycorax, we had more data, better distributed in time than for Caliban,
but relatively to the errors the photometric variation was smaller.
The first night Sycorax displayed
 $\Delta R = 0.076 \pm 0.027$\ mag,
 while in the second one
 $\Delta R = 0.067 \pm 0.026$ mag,
both below $3\sigma$ level. So we cannot claim for a safe positive detection
of significant brightness variation in the Sycorax data, and any attempt to
estimate a period would be considered as tentative, although during the second
night the data hinted at a possible {\em sinusoidal} variation. Indeed, the
hypothesis of random fluctuation over a time constant brightness as an explanation
of the observed light curve, fits Sycorax
data better than Caliban ones, but the significativity level for such a fit is
only $55.6\%$, making this hypothesis hard to support.
Moreover, the R magnitudes for the reference stars in the
Sycorax frames for the first night are stable, with an RMS better than
$0.005$\ mag and a significativity level for the hypothesis of a
constant magnitude better than 99.99\%. Both these observations support
(but do not proof) the hypothesis that
the detected variability in Sycorax data
could be physically significant.
So we applied the sinusoidal fit shown in Fig. 2.

In the case of Sycorax a not null shift $\Delta = (-2.6 \pm
1.3) \times 10^{-2}$\ mag shall be allowed in order to have
a good fit, which is consistent
with the average photometrical error of the tabulated measures
$\approx 0.027$\ mag. On the contrary the imposition of
$\Delta \equiv 0$\ reduces the significativity level for the fit
of about a factor three. Apart from
the method described here, we attempted different ways to get rid of
this shift, all producing similar results about the estimate of the period,
phase and amplitude for the light curve. Our conclusion is that the
shift must be considered as a relevant parameter in what regard the minimization of
$\chi^2$\ only. Of course, since it affects in the same
way all the data taken in the same night, the shift is not
relevant for the colour determination, since colours are obtained
by neighbor data.

Taking the data of both nights into account, the best fit was obtained for
 $P = 4.1156 \pm 0.0416$\ hours,
 $\phi = 4.8750 \pm 0.2594$\ radians,
 $A = 0.0308 \pm 0.0084$\ mag,
 $R_0 = 20.4566 \pm 0.0103$\ mag
 with a $\chi^2 = 4.9796$\ and
 $SL = 97.6\%$.
This represents our best guess for the Sycorax period.
The repetition of the fit using the second night data only
gives instead
 $P = 3.6841 \pm 0.0406$\ hours,
 $\phi = 0.1679 \pm 0.3303$\ radians,
 $A = 0.0320 \pm 0.0083$\ mag,
 with a $\chi^2 = 2.3487$\ and
 $SL = 88.5\%$.
Note that $A$\ is not affected by the change which instead
affects period and phase. In addition the significativity is
higher when all the data are used to perform the fit,
suggesting that all the data have the same statistical significance.
At last $R_0$\ from the second night data only is:
 $R_0 = 20.4322 \pm 0.0053$\ mag whose difference from the $R_0$\ obtained
from the full night is dominated by $\Delta$. After correction of this shift
and adding in square the two random errors we obtain:
$R_0 = 20.4062 \pm 0.0140$\ mag.
In conclusion for Sycorax:
 $P    = 4.12 ^{+0.00}_{-0.43} \pm 0.04$\ hours,
 $\phi = 4.88 ^{+0.00}_{-4.71} \pm 0.25$\ radians,
 $A    = 0.032 ^{+0.001}_{-0.000} \pm 0.008$\ mag,
 $R_0  = 20.457 ^{+0.000}_{-0.001} \pm 0.010$\ mag.

\section{Conclusions}

In the nights between October 8 and 9, 1999 we  carried out
accurate multicolor observations of Uranus' irregular satellites
Sycorax (S/1997 U1) and Caliban (S/1997 U2), providing magnitudes in B, V, R, I
bands.

The colors we obtained confirm the values suggested by
Gladman et al. \cite{Gladman:etal:1998}.
They are redder than Uranus and its
regular satellites, and Sycorax appears to be bluer than Caliban
and most of the Kuiper Belt objects.

We obtained light-curves in the R band for both satellites,
and we estimated periods and amplitudes by fitting the data with a
sinusoid. Caliban's light curve displayed
significant fluctuations (5$\sigma$), which were not evident in the Sycorax data.

For Caliban, we suggest
a light curve period of about $2.7$\ hour with an amplitude of about $0.13$\ mag, which is
compatible with the rotation periods of the
Kuiper-belt objects (Romanishin \& Tegler \cite{Romanishin:Tegler:1999}).
However, the limited number of points and time coverage coupled with calibration
uncertainties, suggest to be conservative and to consider this result just as a first
estimate requiring further observations to be confirmed.

Although the data for Sycorax did not show a so large photometric variation,
we tentatively provide an estimate of the light-curve period and amplitude,
which also has to be considered  preliminary, amounting at
about $3.7 \div 4.1$\ hours and $0.03$\ mag respectively.
Better time coverage together with very accurate photometry
may help to unravel a safer light-curve period for Sycorax.

\section{Acknowledgements}
The authors wish to acknowledge the unknown referee for valuable suggestions and comments.


\clearpage



\figcaption[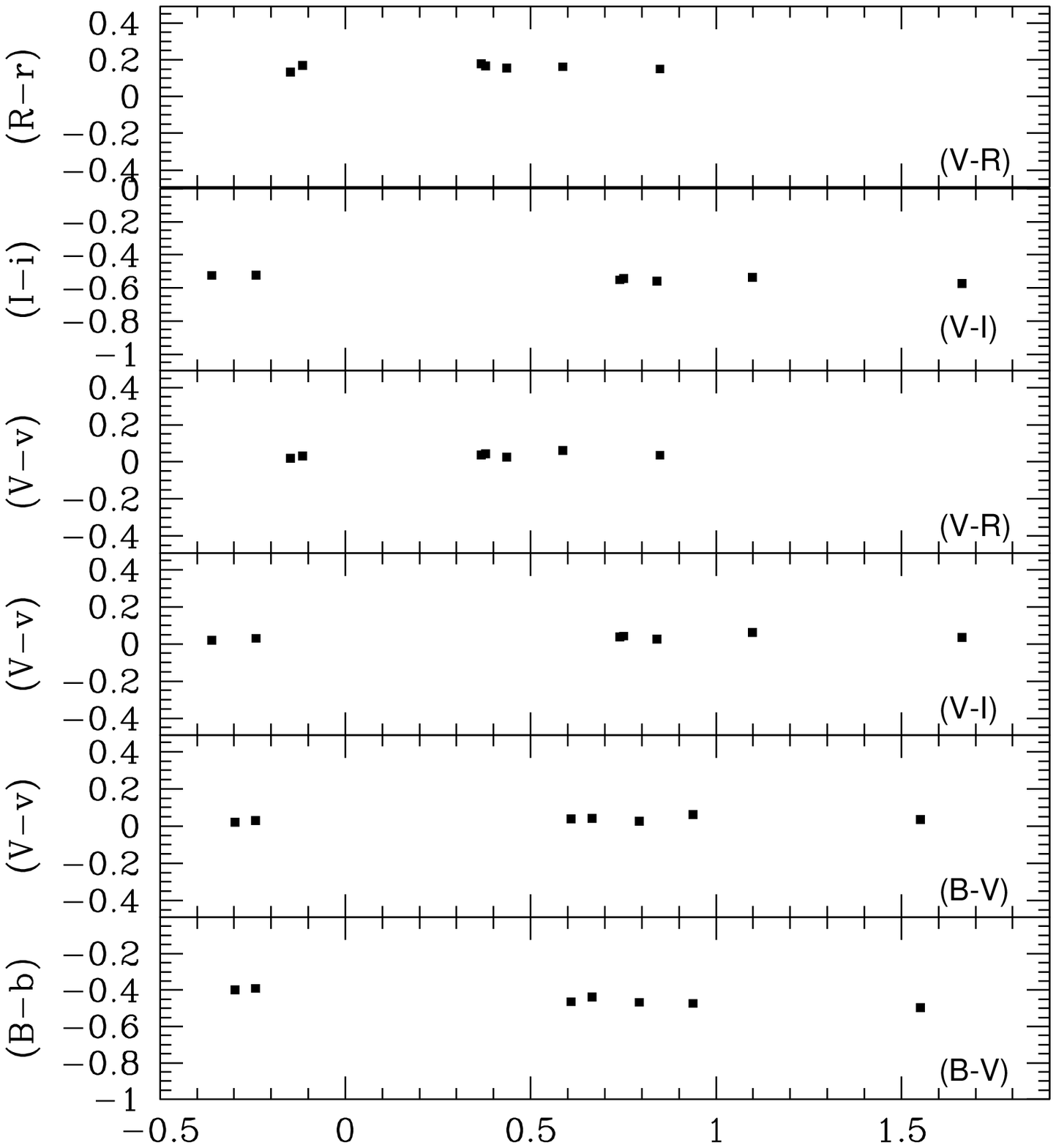]{Colour Equations for the night October
8, 1999\label{fig1}}

\figcaption[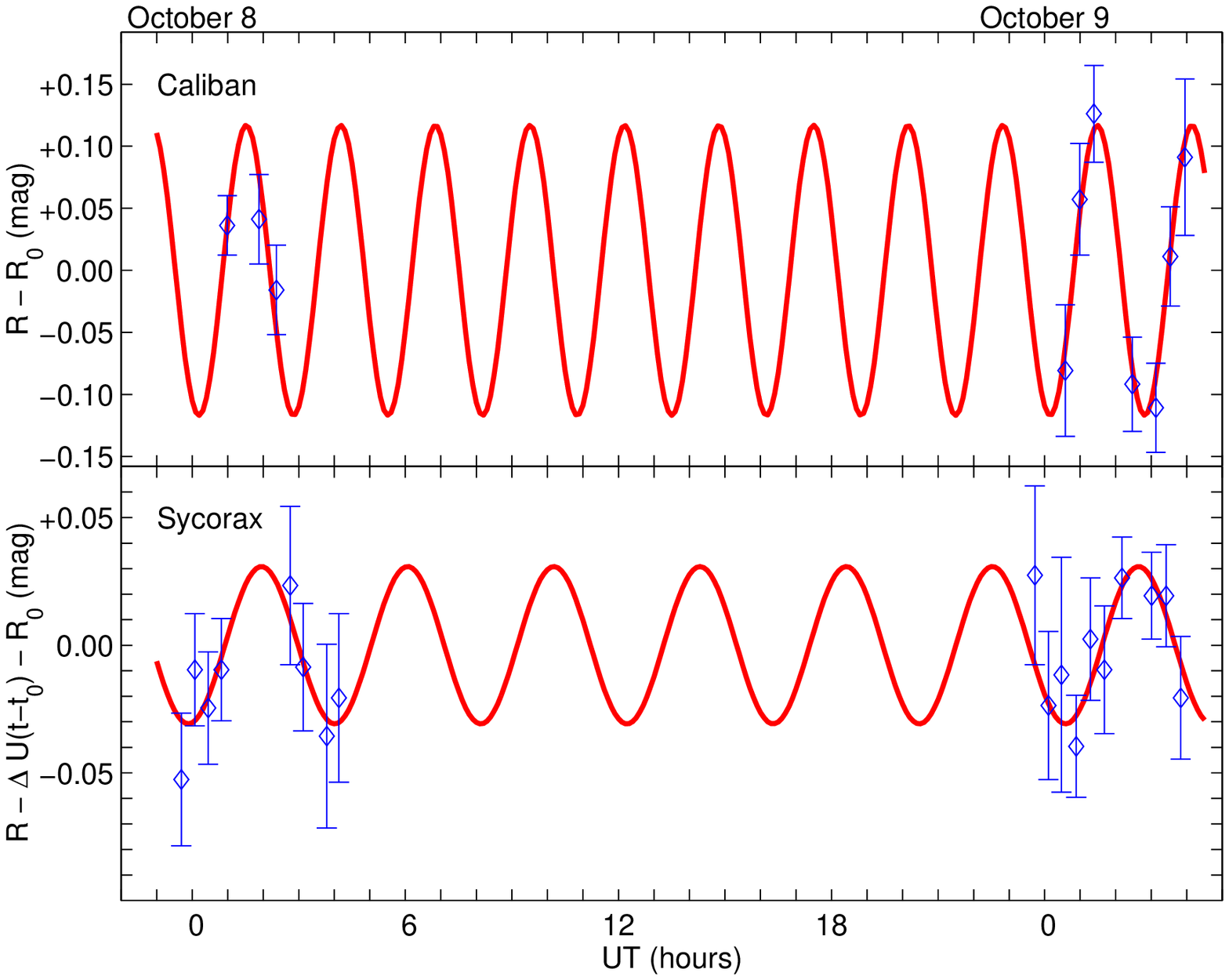]{Caliban and Sycorax lightcurves.
The non-sinusoidal terms of eq. (\ref{eq:sinfit}) have been subtracted
from both the lightcurves to evidence the time variability.
Therefore $R_0$\ has been removed from the Caliban's lightcurve,
while $R_0 + \Delta \, \mathrm{U}(t - t_0) $\ has been removed
from the Sycorax lightcurve.\label{fig2}}

\clearpage
 \begin{table*}
 \tabcolsep 0.095truecm
 \caption{Log of the observations for Sycorax.}\label{tab:Sycorax:Journal}
 \begin{center}
 \begin{tabular}{cccccc} \hline
 \multicolumn{1}{c}{Date} & \multicolumn{1}{c}{Filter} &
 \multicolumn{1}{c}{Exp. time}& \multicolumn{1}{c}{U.T.} &
 \multicolumn{1}{c}{airmass}& magnitude\\
 & &  (sec) & hh:mm:ss.sss&&\\
 &&&&&\\
 October 8, 1999 & R &  80& 23:41:37.210&1.052&20.404$\pm$0.026\\
                & R &   80& 00:04:30.130&1.033&20.447$\pm$0.022\\
                & R &   70& 00:27:03.560&1.023&20.432$\pm$0.022\\
                & B &  300& 00:31:39.030&1.022&21.764$\pm$0.029\\
                & V &   80& 00:39:10.840&1.021&20.752$\pm$0.025\\
                & I &   80& 00:44:14.550&1.021&19.815$\pm$0.021\\
                & R &   70& 00:49:21.380&1.022&20.447$\pm$0.020\\
                & R &   70& 02:45:36.780&1.161&20.480$\pm$0.031\\
                & B &  300& 02:50:14.170&1.173&21.812$\pm$0.030\\
                & V &   80& 02:57:38.700&1.193&20.800$\pm$0.022\\
                & I &   80& 03:02:37.590&1.208&19.863$\pm$0.021\\
                & R &   70& 03:07:28.890&1.223&20.448$\pm$0.025\\
                & R &   80& 03:47:31.370&1.386&20.421$\pm$0.036\\
                & R &   80& 04:07:50.470&1.501&20.436$\pm$0.033\\
October 9, 1999 & R &  120& 23:43:27.090&1.047&20.458$\pm$0.035\\
                & R &  120& 00:06:17.240&1.030&20.407$\pm$0.029\\
                & R &  120& 00:28:17.050&1.022&20.419$\pm$0.046\\
                & R &  120& 00:53:16.650&1.023&20.391$\pm$0.020\\
                & R &  120& 01:17:19.370&1.034&20.433$\pm$0.024\\
                & R &  200& 01:40:47.210&1.054&20.471$\pm$0.025\\
                & B &  500& 01:47:33.150&1.062&21.915$\pm$0.031\\
                & V &  150& 01:58:24.400&1.076&20.903$\pm$0.025\\
                & I &  150& 02:04:32.810&1.086&19.966$\pm$0.022\\
                & R &  200& 02:10:43.380&1.096&20.457$\pm$0.016\\
                & R &  200& 03:00:34.960&1.214&20.450$\pm$0.017\\
                & R &  200& 03:25:09.110&1.302&20.450$\pm$0.020\\
                & R &  200& 03:49:51.190&1.419&20.410$\pm$0.024\\
 \hline
 \end{tabular}
 \end{center}
 \end{table*}

\clearpage

 \begin{table*}
 \tabcolsep 0.10truecm
 \caption{Log of the observations for Caliban.}\label{tab:Caliban:Journal}
 \begin{center}
 \begin{tabular}{cccccc} \hline
 \multicolumn{1}{c}{Date} &
 \multicolumn{1}{c}{Filter} &
 \multicolumn{1}{c}{Exp. time}&
 \multicolumn{1}{c}{U.T.} &
 \multicolumn{1}{c}{airmass}&
 magnitude\\
 & &  (sec) & hh:mm:ss.sss&\\
 &&&&&\\
 October 8, 1999& R &  600& 00:59:10.500&1.024&21.949$\pm$0.024\\
                & B & 1200& 01:22:37.740&1.029&23.659$\pm$0.049\\
                & V &  800& 01:35:08.840&1.046&22.423$\pm$0.033\\
                & R &  600& 01:52:00.450&1.064&21.954$\pm$0.036\\
                & I &  800& 02:05:33.450&1.083&21.440$\pm$0.024\\
                & R &  600& 02:22:24.310&1.113&21.897$\pm$0.036\\
 October 9, 1999& R &  800& 00:34:52.990&1.021&21.832$\pm$0.053\\
                & R &  800& 00:59:19.070&1.025&21.970$\pm$0.045\\
                & R &  800& 01:23:16.540&1.039&22.039$\pm$0.039\\
                & R &  800& 02:18:08.040&1.112&21.821$\pm$0.038\\
                & R &  800& 03:07:55.570&1.243&21.802$\pm$0.036\\
                & R &  800& 03:32:22.580&1.339&21.924$\pm$0.040\\
                & R &  800& 03:56:58.850&1.468&22.004$\pm$0.063\\
\hline
\end{tabular}
\end{center}
\end{table*}

 \begin{table*}[t]
 \caption{Colors of Sycorax and Caliban}\label{tab:colours}
 \begin{center}
 \begin{tabular}{cccccc}
 \hline
  Satellite  &  B-V  &  V-R  &  R-I & B-R & V-I \\
 \hline
 Sycorax  &  1.012$\pm$0.038  &  0.482$\pm$0.042 & 0.455$\pm$0.030 & 1.494$\pm$0.036 & 0.937$\pm$0.031\\
 Caliban  &  1.236$\pm$0.059  &  0.473$\pm$0.048 & 0.510$\pm$0.043 & 1.709$\pm$0.054 & 0.983$\pm$0.041\\
 \hline
 \end{tabular}
 \end{center}
 \end{table*}

\clearpage

\begin{figure}
\plotone{maris.fig1.eps}
\end{figure}

\begin{figure}
\plotone{maris.fig2.eps}
\end{figure}

\end{document}